\documentclass[12pt]{article}

\usepackage{amsmath}
\usepackage{amssymb}
\usepackage{amsfonts}
\usepackage{lscape}

\begin{document}

\fontsize{12}{6mm}\selectfont
\setlength{\baselineskip}{2em}

$~$\\[.35in]
\newcommand{\dss}{\displaystyle}
\newcommand{\raro}{\rightarrow}
\newcommand{\be}{\begin{equation}}

\def\sech{\mbox{\rm sech}}
\def\sn{\mbox{\rm sn}}
\def\dn{\mbox{\rm dn}}
\thispagestyle{empty}

\begin{center}
{\Large\bf A Time Dependent Version of  } \\    [2mm]
{\Large\bf the Quantum WKB Approximation } \\    [2mm]
\end{center}

\vspace{1cm}
\begin{center}
{\bf Paul Bracken}                        \\
{\bf Department of Mathematics,} \\
{\bf University of Texas,} \\
{\bf Edinburg, TX  }  \\
{78541-2999}
\end{center}

\vspace{3cm}
\begin{abstract}
The phenomenon of quantum tunneling is reviewed and an 
overview of applying approximate methods for studying
this effect is given. An approach to a time-dependent
formalism is proposed in one dimension and generalized to higher
dimensions. Some physical examples
involving the resulting wavefunction which is determined
are presented.
\end{abstract}

\vspace{2mm}

\vspace{2mm}
Keywords: wavefunction, tunneling, time-dependent, Hamiltonian

\vspace{2mm}
PACS: 31.15.Gy, 31.15.Md

\newpage
There is a great deal of interest in integrable systems
at the moment and their quantization. The semiclassical treatment
of integrable systems has its roots in Bohr's atomic model as well
as in Einstein's well known paper on the quantization of regular motion {\bf [1]}.
It is known that semiclassical quantization is exact only
for so-called solvable potentials, which would 
include the harmonic oscillator {\bf [2,3]}. The nonsolvable
case yields approximate results which may be asymptotic in 
nature {\bf [4,5]}. Of course there are summation techniques which
can be applied to divergent series. Recently {\bf [6]} a
quantum phase was derived which gives exact quantization
of nonsolvable potentials and avoids many of these 
complications. A consistent divergence-free scheme is
still an active area of research.

The phenomenon of tunneling has been a characteristic
property of many types of quantum mechanical processes,
especially those which involve confining potential wells {\bf [7]}.
Their mathematical description has generated many
different types of results. Due to the wide variety of
applied applications where tunneling plays some role,
the study of mathematical methods which would be
useful in this context is still very active. 
Along these lines, the quantum WKB approximation is
very important, and not a great deal exists in the literature for 
the time-dependent case {\bf [8]}. An important example
of a system where this approximation can be useful
is the case of the decay of a system from a long-lived
metastable state which frequently arises in physics.
There has always been considerable interest in
understanding the quantum tunneling of a macroscopic
coordinate. If the height of a potential barrier is greater
than the total energy $E$ of a particle, the kinetic 
energy of the particle is negative and so the study of 
the tunneling of a particle appears to have a paradoxical 
aspect to it. However this approach implies that we can
assign values to the coordinate $x$ and the momentum
$p$ simultaneously {\bf [7]}. This however would violate the
uncertainty principle. If however for a very short time
$\Delta t$, the uncertainty in the energy $\Delta E$ 
is such that the total energy of the particle is
greater than the height of the barrier, then
tunneling takes place in time $\Delta t$ if the 
particle can traverse the barrier in this time interval.
The continued interest in the Josephson effect,
and the fact that high precision measurements are
possible for such systems probably accounts for
some of this interest. Another reason is that decay
rates can be calculated theoretically based on
models for such systems. The purpose here is to
begin to introduce the treatment of time dependence
into these applications, and to discuss some of their
quantum consequences. A perturbative time-dependent 
wavefunction will be developed for the time-dependent
Schr\"{o}dinger equation, and some applications will
be considered. In particular, this type
of wave function is applicable to the study of the Berry phase.

The time-dependent Schr\"{o}dinger equation in three
space dimensions has the form
$$
i \hbar \frac{\partial \psi}{\partial t} = H \psi,
\eqno(1)
$$
where $H$ is the Hamiltonian for the system.
The case in which the Hamiltonian contains a
potential function will be of interest here,
in which case (1) can be written
$$
i \hbar \frac{\partial \psi}{\partial t}
= - \frac{\hbar^2}{2m} \Delta \psi + V \psi.
\eqno(2)
$$
Here $\Delta$ is the Laplacian operator in terms of
the position coordinates. To motivate the approach,
consider a function $\psi$ which is to satisfy (2)
which has the particular form
$$
\psi ( {\bf x}, t) = A ({\bf x}, t)
\displaystyle e^{\frac{i}{\hbar} S ( {\bf x}, t)}.
\eqno(3)
$$
In (3), both $A$ and $S$ depend on both the
spatial variables and the time variable. 
Substituting (3) into (2), after differentiating
with respect to the given variables, (2) takes
the form 
$$
i \hbar \frac{\partial A}{\partial t}
- A \frac{\partial S}{\partial t}
= - \frac{\hbar^2}{2m}
[  \Delta A + 2 \frac{i}{\hbar} 
{\bf \nabla} A \cdot {\bf \nabla} S
+ \frac{i}{\hbar} A \Delta S -
\frac{1}{\hbar^2} A ( {\bf \nabla} S)^2 ] + VA.
\eqno(4)
$$
Equating the real and imaginary parts on both
sides of (4), two independent, coupled equations
in terms of the variables $A$ and $S$ then result
and are given by
$$
\frac{\partial S}{\partial t} + \frac{1}{2m} ( {\bf \nabla} S)^2
+ V - \frac{\hbar^2}{2m A} \Delta A =0,
\eqno(5)
$$
$$
\frac{\partial A}{\partial t} + \frac{A}{2m} \Delta S
+ \frac{1}{m} {\bf \nabla} S \cdot {\bf \nabla } A = 0.
\eqno(6)
$$
Of course, by neglecting the term proportional to $\hbar^2$
in (5), the classical Hamilton-Jacobi equation in
terms of the single function $S$ results. In this approximation,
a solution to (5) in (6) gives an
equation for the function $A$.

Let us now consider another version of this
process. In other words, let $\psi$ have the basic
structure given in (3), and then require that $S$ satisfy 
the classical Hamilton-Jacobi equation. In fact, it
proves more appropriate to introduce a function which is 
a superposition of such functions in powers of $\hbar$.

Consider a classical system which is governed by a
Hamiltonian $H$, then the associated classical Hamilton-Jacobi
equation has the following form in terms of
$r$-space variables
$$
\frac{\partial S}{\partial t} + H ( q_1, \cdots, q_r,
\frac{\partial S}{\partial q_1}, \cdots,
\frac{\partial S}{\partial q_r}, t) = 0.
\eqno(7)
$$
Although it should be possible to generalize
what follows to higher dimensions, let us suppose
that there is just one spatial variable in order
to simplify the discussion. Then for a Hamiltonian 
of the form that generates (2), equation (7)
takes the form
$$
\frac{1}{2m} ( S_x)^2 + V + \frac{\partial S}{\partial t} =0.
\eqno(8)
$$
The Schr\"odinger equation (2) in one variable
reduces to the form
$$
( \frac{\hbar^2}{2m} \partial_x^2 - V + i \hbar \frac{\partial}{\partial t})
\psi = 0.
\eqno(9)
$$
Define the linear operator
$$
L = \frac{\hbar^2}{2m} \partial_x^2 - V + i \hbar \frac{\partial}{\partial t}.
\eqno(10)
$$
Equation (9) then takes the compact form
$$
L \psi =0.
\eqno(11)
$$
Now the wave function $\psi$ in (3) will be generalized
to the following series form in ascending powers of $\hbar$
by writing
$$
\Psi (x,t) = \sum_{k=0}^N (i \hbar)^k a_k (x,t) \,
\exp (\frac{i}{\hbar} S (x,t)).
\eqno(12)
$$
In expression (12), the functions $S (x,t)$
and the coefficients $a_k (x,t)$ depend on both the
space and time variables. Moreover, the $a_k$ are to be
real-valued $a_k : \mathbb R \rightarrow \mathbb R$.
It will be seen that it is possible to force (12) to
satisfy (11) up to a single error term which is
proportional to a power of $\hbar$ by imposing a
set of differential constraint equations on the
coefficients $a_k$ which appear in (12).

The second derivative of $\Psi$ is required in (9)
and it is given by
$$
\frac{\hbar^2}{2m} \Psi_{xx} = \frac{1}{2m} \{ -i \hbar
\sum_{k=1}^{N} (i \hbar)^k a_{k-1,xx} + 2 i \hbar S_x
\sum_{k=0}^N ( i \hbar)^k a_{k,x} - (S_x)^2
\sum_{k=0}^{N} (i \hbar)^k a_k
$$
$$
- (i \hbar)^{N+2} a_{N,xx} + i \hbar S_{xx}
\sum_{k=0}^{N} (i \hbar)^k a_k \} \exp ( i \frac{S}{\hbar}).
\eqno(13)
$$
The time derivative is given by
$$
i \hbar \Psi_t = \sum_{k=0}^N ( i \hbar )^{k+1} a_{k,t}
\exp ( i \frac{S}{\hbar}) - S_t \Psi.
\eqno(14)
$$
Here, the variable subscript indicates differentiation 
with respect to $x$ or $t$. Substituting these
derivatives into Schr\"odinger equation (9), we obtain
$$
\frac{\hbar^2}{2m} \Psi_{xx} - V \Psi + i \hbar \Psi_t
= \frac{1}{2m} i \hbar \exp ( \frac{i}{\hbar} S) \sum_{k=1}^{N} \{ - (i \hbar)^k a_{k-1,xx} 
+ 2 S_x (i \hbar)^k a_{k,x} + S_{xx} (i \hbar)^k a_k \} 
$$
$$
- \frac{1}{2m} (i \hbar)^{N+2} a_{N,xx} \, \exp( \frac{i}{\hbar} S)
+ \frac{1}{2m} i \hbar \{ 2 S_x a_{0,x} + S_{xx} a_{0} \} 
\exp( \frac{i}{\hbar} S)
\eqno(15)
$$
$$
- ( \frac{1}{2m} S_{x}^2 \Psi + V \Psi + S_{t} \Psi )
+ \sum_{k=0}^N ( i \hbar)^{k+1} a_{k,t} \exp ( \frac{i}{\hbar} S).
$$
Now suppose that the function $S (x,t)$ satisfies
the Hamilton-Jacobi equation (8), then (15) takes
the form
$$
L \Psi = \frac{1}{2m} \, \exp (\frac{i}{\hbar} S)
\{ \sum_{k=1}^N ( i \hbar)^{k+1} ( - a_{k-1,xx} + 2 S_x a_{k,x}
+ S_{xx} a_{k} + 2m a_{k,t} ) \}
$$
$$
+ \frac{1}{2m} i \hbar \exp ( \frac{i}{\hbar} S)
\{ 2 S_x a_{0,x} + S_{xx} a_0 + 2m a_{0,t} \}
- \frac{1}{2m} ( i \hbar)^{N+2} a_{N,xx} \exp ( \frac{i}{\hbar} S).
\eqno(16)
$$
It is now required that for a given function $S$, the
coefficient functions $a_k$ must satisfy the following
system of first-order 
partial differential equations
with the definition $a_{-1} =0$,
$$
2 S_x a_{0,x} + S_{xx} a_{0} + 2 m a_{0,t} = 0,
\eqno(17)
$$
$$
2 S_x a_{k,x} + S_{xx} a_{k} + 2 m a_{k,t} - a_{k-1,xx}=0,
\quad
k=1, \cdots, N.
\eqno(18)
$$
Equation (18) is non-homogeneous since the coefficient $a_{k-1}$
and hence $a_{k-1,xx}$ 
has been determined from (17)-(18) at the previous order $k-1$.
In the event that (17) and (18) hold, all the terms
in the brackets on the right-hand side of (16) vanish,
and the Schr\"odinger equation reduces to simply
$$
L \Psi = - \frac{1}{2m} (i \hbar )^{N+2} a_{N,xx} \exp 
( \frac{i}{\hbar} S).
\eqno(19)
$$
If wavefunction (12) had been an exact solution of
(9), the right-hand side of (19) would have reduced to zero.
Thus this procedure in which $\Psi$ is taken in the form (12)
provides a solution of (9) up to an error term 
proportional to $\hbar^{N+2}$, provided that $S$ is a
solution of the classical Hamilton-Jacobi equation (8),
and the $a_k$ satisfy the set of partial differential
equations (17)-(18). Upon letting $N \rightarrow \infty$,
an asymptotic solution to equation (11) results which
has the form
$$
\Psi \sim ( \sum_{k=0}^{\infty} a_k (x,t) \hbar^k )
\exp ( \frac{i}{\hbar} S).
\eqno(20)
$$
Once $S$ has been 
determined, the coefficients $a_k$ are obtained
recursively as solutions of equations (17)-(18)
which generate a time dependent solution.
Convergence of (20) will not be discussed here.

This procedure can be generalized to higher spatial
dimensions by using the linearity of the Schr\"odinger 
equation (2). The Hamilton-Jacobi equation has the form
$$
\partial_t S + \frac{1}{2m} ( {\bf \nabla} S )^2 + V =0.
\eqno(21)
$$
Thus, if the wavefunction $\Psi$ depends on three spacial
coordinates ${\bf x}$, the analogue of (12) will be written
$$
\Psi ( {\bf x}, t) = \sum_{k=0}^{N} ( i \hbar)^k 
a_k ( {\bf x}, t) \exp ( \frac{i}{\hbar} S ( {\bf x}, t)).
\eqno(22)
$$
The $\Delta \Psi$ term in (2) will include three
expressions of the form (13) for each of the variables.
Therefore, we can write
$$
L \Psi = \frac{\hbar^2}{2m} \Delta \Psi - V \Psi
+ i \hbar \partial_t \Psi
$$
$$
= \frac{1}{2m} i \hbar \exp ( \frac{i}{\hbar} S)
\sum_{k=1}^N [ - ( i \hbar)^k \Delta a_{k-1} + 2 (i \hbar)^k
( {\bf \nabla} S ) \cdot
{\bf \nabla} a_k + (i \hbar)^k a_k \Delta S ]
$$
$$
- (i \hbar)^{N+2} \Delta a_n \exp (\frac{i}{\hbar} S) 
+ \frac{i}{2m} \hbar [ 2 ( {\bf \nabla} S) \cdot ({\bf \nabla } a_0)
+ \Delta S \, a_0 ] \exp( \frac{i}{\hbar} S)
\eqno(23)
$$
$$
- ( \frac{1}{2m} ( {\bf \nabla} S)^2 + V + \partial_t S) \Psi
+ \sum_{k=0}^N (i \hbar)^{k+1}
a_{k,t} \exp( \frac{i}{\hbar} S).
$$
If we suppose that $S$ is a function which
satisfies the classical Hamilton-Jacobi equation (21),
then the coefficient functions $a_k$ are required to
satisfy the following system analogous to the pair (17) and (18)
$$
\begin{array}{c} 
2 {\bf \nabla} S \cdot {\bf \nabla} a_0 +
\Delta S a_0 + 2 m a_{0,t} =0,    \\
    \\
2 {\bf \nabla} S \cdot {\bf \nabla} a_k +
\Delta S a_k + 2m a_{k,t} - \Delta a_{k-1} =0,
\quad
k=1, \cdots, N    \\
\end{array}
\eqno(24)
$$
Under the constraints (21) and (24), (23) reduces to
$$
L \Psi = - \frac{1}{2m} ( i \hbar)^{N+2} \Delta a_N \,
\exp( \frac{i}{\hbar} S).
\eqno(25)
$$

Consider as an example the realistic
case of the one-dimensional harmonic oscillator Hamiltonian
given by
$$
H = \frac{1}{2m} p^2 + x^2.
\eqno(26)
$$
Then it is easy to check that the function
$$
S = \int \sqrt{2m ( \beta - x^2)} \, dx - \beta t
\eqno(27)
$$
is a solution to (8) under the harmonic 
potential function. In (27), $\beta$ is a 
separation constant obtained upon integrating (8).
Substituting (27) into (17), $a_0$ will be given
by the first order equation
$$
2 \sqrt{2m} \sqrt{\beta - x^2} \frac{\partial a_0}{\partial x}
- \sqrt{2m} \frac{x}{\sqrt{\beta - x^2}} a_0 + 2m
\frac{\partial a_0}{\partial t} = 0.
\eqno(28)
$$
This can be solved by the method of characteristics and the
general solution is given by
$$
a_{0} (x,t) = ( \beta - x^2)^{1/4} \varphi ( t - \frac{\sqrt{2m}}{2}
\arctan ( \frac{x}{\sqrt{\beta - x^2}})).
\eqno(29)
$$
Using (27), $a_0$ in (28) can be written in terms
of the derivative of $S$ as
$$
a_0 (x,t) = ( \frac{S_x}{\sqrt{2m}})^{1/2} \varphi ( t - 
\frac{\sqrt{2m}}{2} \arctan ( \frac{\sqrt{2m} x}{S_x} ).
$$
In (29), $\varphi$ is selected to be an arbitrary twice differentiable
function of a single variable. 
The method of characteristics does not specify how $\varphi$
is to be chosen, however, this can be done on physical grounds.
Higher order coefficients
for the expansion (12) can be obtained from (18)
using (29), however they are rather more complicated
and will not be presented here.

As a final application of wavefunction (22), let us 
consider a curve $C$ on a manifold of external parameters
$M$, and the adiabatic evolution of the quantum system 
described by the parameter dependent Hamiltonian $H = H({\bf v})$
along the curve $C$, and ${\bf v}$ represents the set of parameters
in vector form.
A one-form may be defined on $M$ as follows
$$
A^{(n)} = - Im \langle n | d_M n \rangle.
$$
The integral of the one-form $A^{(n)}$ along a curve $C$
between ${\bf v}_0$ and ${\bf v}_1$ in parameter space produces the
following geometric quantity
$$
\gamma_n (C) = \oint_C A^{(n)} = \int_{\Sigma} F^{(n)}.
\eqno(30)
$$
This is the Berry phase,
where $\Sigma$ is an arbitrary two-dimensional
submanifold of $M$ such that $\partial \Sigma =C$ and
$F^{(n)} = d A^{(n)}$. In terms of the wave function
$\psi_n ({\bf x}; {\bf v})$, the two-form is calculated from
$A^{(n)}$ to be
$$
F^{(n)} = - Im [ d_M \int d^N x \,
\psi^{*} ( {\bf x} ;{\bf v}) \, d_M \, \psi_n ({\bf x}; {\bf v}) ].
$$
There is much interest in the study of the
connection between the quantal Berry phase $\gamma_n (C)$ 
and the classical Hannay angle. 
An important question in this regard asks if a
classical system has a Hannay angle,
will it also possess a Berry phase when it is quantized.
Wavefunctions of the
type discussed here
can be useful for discussing questions regarding the
relationship bewteen these quantities.

\begin{center}
{\bf References.}
\end{center}

\noindent
$[1]$ A. Einstein, Verh. Dtsch. Phys. Ges. {\bf 19}, 82 (1917),
A. Stone, Phys. Tod. {\bf 8}, 57 (2005).  \\
$[2]$ J. B. Keller, Ann. Phys. {\bf 4}, 180 (1958).  \\
$[3]$ J. B. Keller, Siam Rev. {\bf 27}, 485 (1985).  \\
$[4]$ T. Hyouguchi, R. Seto, M. Ueda and S. Adachi,
Ann. Phys. {\bf 312}, 177 (2004).  \\
$[5]$ C. M. Bender, K. Olaussen and P. S. Wang, Phys. Rev.
{\bf D 16}, 1740 (1977).  \\
$[6]$ M. Matzkin, Phys. Rev. {\bf A72}, 054102 (2005).  \\
$[7]$ Quantum Theory of Tunneling, M. Razavy, World
Scientific (2003).   \\
$[8]$ M. P. A. Fisher, Phys. Rev. {\bf B 37}, 75 (1988).  \\

\end{document}